\documentclass[a4paper,12pt]{article}

\usepackage{amsmath}
\usepackage{amssymb}

\makeatletter
\def\presectname{}
\def\postsectname{}
\def\presubsectname{}
\def\postsubsectname{}
\def\presubsubsectname{}
\def\postsubsubsectname{}
\def\preappendixname{}
\def\postappendixname{}
\def\presubappendixname{}
\def\postsubappendixname{}
\def\presubsubappendixname{}
\def\postsubsubappendixname{}
\renewcommand\section{
  \def\@presectname{\presectname}
  \def\@postsectname{\postsectname}
  \@startsection {section}{1}{\z@}%
                                   {-3.5ex \@plus -1ex \@minus -.2ex}%
                                   {2.3ex \@plus.2ex}%
                                   {\normalfont\Large\bfseries}}
\renewcommand\subsection{
  \def\@presectname{\presubsectname}
  \def\@postsectname{\postsubsectname}
  \@startsection{subsection}{2}{\z@}%
                                     {-3.25ex\@plus -1ex \@minus -.2ex}%
                                     {1.5ex \@plus .2ex}%
                                     {\normalfont\large\bfseries}}
\renewcommand\subsubsection{
  \def\@presectname{\presubsubsectname}
  \def\@postsectname{\postsubsubsectname}
  \@startsection{subsubsection}{3}{\z@}%
                                     {-3.25ex\@plus -1ex \@minus -.2ex}%
                                     {1.5ex \@plus .2ex}%
                                     {\normalfont\normalsize\bfseries}}
\def\@seccntformat#1{\@presectname\csname the#1\endcsname\@postsectname\quad}
\@addtoreset{equation}{section}
\renewcommand{\theequation}{\thesection.\@arabic\c@equation}
\renewcommand\appendix{\par\newpage
  \setcounter{section}{0}%
  \setcounter{subsection}{0}%
  \gdef\thesection{\@Alph\c@section}
  \renewcommand{\presectname}{\preappendixname}
  \renewcommand{\postsectname}{\postappendixname}
  \renewcommand{\presubsectname}{\presubappendixname}
  \renewcommand{\postsubsectname}{\postsubappendixname}
  \renewcommand{\presubsubsectname}{\presubsubappendixname}
  \renewcommand{\postsubsubsectname}{\postsubsubappendixname}}
\makeatother

\newcommand{\ssum}[2]{\mathop{\overset{#2}{\underset{#1}{\textstyle\sum}}}}

\newcommand{\tr}{\mathop{\mathrm{tr}}}
\newcommand{\e}{\mathrm{e}}
\newcommand{\rmd}{\mathrm d}

\newcommand{\Repa}{\mathop{\mathrm{Re}}}
\newcommand{\Impa}{\mathop{\mathrm{Im}}}

\newcommand{\deriv}[2]{\frac{\partial #1}{\partial #2}}

\newcommand{\bracket}[1]{\left\langle #1\right\rangle}

\newcommand{\Const}{\text{Const}}
\newcommand{\eff}{\text{eff}}

\renewcommand{\preappendixname}{\appendixname\ }
\renewcommand{\postappendixname}{:\hspace{-10pt}}

\begin{document}

\titlepage

\vspace*{-15mm}   
\baselineskip 10pt   
\begin{flushright}   
\begin{tabular}{l}   
{December 2004} \\ 
{KUNS-1946}\\
{RIKEN-TH-33}\\
{hep-th/0412004}   
\end{tabular}   
\end{flushright}   
\baselineskip 24pt   
\vglue 10mm   

\begin{center}
{\Large\bf
 Universality of Nonperturbative Effect \\
 in Type 0 String Theory
}

\vspace{8mm}   

\baselineskip 18pt   

Hikaru Kawai$^{1,2}$,
Tsunehide Kuroki$^2$ and
Yoshinori Matsuo$^1$ 

\vspace{5mm}   

{\it  
$^1$ Department of Physics, Kyoto University, 
     Kyoto 606-8502, Japan 
\\ 
$^2$ Theoretical Physics Laboratory, RIKEN, 
     2-1 Wako, Saitama 351-0198, Japan  
}
  
\vspace{10mm}   

\end{center}

\begin{abstract}
\noindent
We derive the nonperturbative effect in type 0B string theory, 
which is defined by taking the double scaling limit of 
a one-matrix model with a two-cut eigenvalue distribution. 
However, the string equation thus derived 
cannot determine the nonperturbative effect completely, 
at least without specifying unknown boundary conditions. 
The nonperturbative contribution to the free energy 
comes from instantons in such models.
We determine by direct computation in the matrix model 
an overall factor of the instanton contribution, 
which cannot be determined by the string equation itself.
We prove that it is universal in the sense 
that it is independent of the detailed structure of potentials 
in the matrix model. 
It turns out to be a purely imaginary number and therefore 
can be interpreted as a quantity related to instability 
of the D-brane in type 0 string theory.  We also comment 
on a relation between our result and boundary conditions 
for the string equation.

\end{abstract}

\baselineskip 18pt   

\newpage

\section{Introduction}\label{sec:intro}

The nonperturbative effect in string theory can be 
studied using matrix models. In particular, 
the noncritical string theory, which is a simplified model of 
string theory, is exactly solvable via matrix models
\cite{Brezin:1990rb}. 
The string equation, which can be derived from them, 
contains the nonperturbative effect of the noncritical string theory
\cite{David:1990sk}. 
On the other hand, study of the Liouville theory 
\cite{Knizhnik:1988ak} enables us 
to find the effect of the D-brane, which is 
the nonperturbative effect of string theory 
and can be identified with the effect that appears 
in the string equation. 
In \cite{Hanada:2004im}, we have shown 
that the string equation does not describe the 
nonperturbative effect completely, at least 
in $c=0$ noncritical string theory. 
To obtain the whole nonperturbative effect, 
it is necessary to study the matrix model directly. 

Recently, 
a matrix model that corresponds to the noncritical string with 
worldsheet supersymmetry has been proposed 
\cite{Douglas:2003up,Klebanov:2003wg}. 
For the $\hat c=0$ noncritical string, 
which is described as two-dimensional pure supergravity 
on the worldsheet \cite{Distler:1989nt}, 
we consider the double scaling limit 
around the Gross-Witten transition \cite{Gross:1980he}. 
This critical point can be found in 
the one-matrix model with a two-cut eigenvalue distribution. 
In string theoretical interpretation, 
the one-matrix model with two cuts corresponds 
to the NSR string theory of type 0B. 
This matrix model can be solved with the string equation. 
However, as in the case of $c=0$ string theory, 
it does not contain the nonperturbative effect completely. 

In this paper we study the nonperturbative effect of 
type 0B string theory by analyzing 
the matrix model with a two-cut eigenvalue distribution. 
We compute the effect of instantons directly in the matrix model, 
which corresponds to the D-brane in the string theory. 
The result is summarized as follows. 

From the string equation, the nonperturbative effect 
in the free energy is obtained as 
\begin{equation}
 \mu = \frac{C}{t^\frac34}
  \exp\left[-\frac{2}{3}t^\frac32\right], 
\end{equation}
where $C$ cannot be determined from the string equation by itself 
without specifying unknown boundary conditions. 
{}From the direct computation using the matrix model, 
we can determine the constant $C$ as 
\begin{equation}
 C = \frac{i}{4\sqrt\pi}.
\end{equation}
Moreover, it is shown that this value is universal, namely, 
it does not depend on the detailed structure in the potential 
of the matrix model. 
Because it is purely imaginary, this nonperturbative effect 
is related to the instability of the D-brane. 

The paper is organized as follows. 
In section~\ref{sec:inst}, we identify the instantons 
in the matrix model and the contribution from instantons 
to the free energy. 
In section~\ref{sec:orth_poly}, we compute the effect of 
the instantons using the method of orthogonal polynomials. 
In section~\ref{sec:universality}, 
we take the double scaling limit and 
consider the universal behavior of the effect of instantons. 
In section~\ref{sec:conclusion}, 
we present the conclusions. 
The appendices show the details of calculations.

\section{Instanton in one-matrix model}\label{sec:inst}

In this section, we consider a one-matrix model. 
We discuss how an instanton contributes to the partition function 
and the free energy. 
The one-matrix model with a one-cut eigenvalue distribution 
corresponds to $c=0$ noncritical string theory, 
while that with a two-cut distribution 
corresponds to $\hat c=0$ type 0B string theory \cite{Douglas:2003up}. 
In both cases, an instanton can be interpreted as 
the ZZ-brane \cite{Zamolodchikov:2001ah}. 
In the one-cut case,  
the instanton can be described as a configuration in which 
all eigenvalues are at the minimum of the effective potential 
except that a single eigenvalue is at its maximum.
This description can be extended to the case of two-cut 
distribution.
In this case, the effective potential has two separated minima. 
Half of the eigenvalues are in one of these minima and 
the other half are in the other minimum except that 
a single eigenvalue is at the maximum. 

In the one-matrix model, the partition function is given by 
\begin{equation}
 Z = \int\rmd\Phi\,\e^{-N\tr V(\Phi)}.
 \label{Z}
\end{equation}
Here, $\Phi$ is an $N\times N$ Hermitian matrix. 
Hereafter we consider the case 
where the potential $V(x)$ is invariant under $x \to -x$ 
and thus the eigenvalue distribution of $\Phi$ 
has this $Z_2$ symmetry.

Diagonalizing the matrix $\Phi$, the partition function can be 
expressed as 
\begin{equation}
 Z = \int\prod_i\rmd\lambda_i\,\Delta^2(\lambda)\,
  \e^{-N\ssum{i}{}V(\lambda_i)}.
\end{equation}
Here, 
$\Delta(\lambda)=\prod_{i<j}\left(\lambda_i-\lambda_j\right)$ 
is the Vandermonde determinant.
We concentrate on the $N$-th eigenvalue $\lambda_N$, 
and represent it as $x$. 
The other $N-1$ eigenvalues can be regarded as those of 
a $(N-1)\times (N-1)$ matrix. 
The partition function of an $N\times N$ matrix model can be expressed 
using an $(N-1)\times (N-1)$ matrix model as
\begin{subequations}\label{EffectivePotential}
 \begin{align}
  Z &= \int\rmd x\int\prod_{i=1}^{N-1}\rmd\lambda_i\,
  \Delta_{N-1}^2(\lambda)\prod_{i=1}^{N-1}(x-\lambda_i)^2
  \,\e^{-N\ssum{i=1}{N-1}V(\lambda_i)-NV(x)}\\
  &= Z_{N-1}\int\rmd x
  \bracket{\det(x-\Phi)^2}_{N-1}
  \e^{-NV(x)}\\
  &= Z_{N-1}\int\rmd x\,\e^{-NV_\eff(x)}.
 \end{align}
\end{subequations}
Here, the subscript ``$N-1$'' indicates that the quantities concerned 
are those in the $(N-1) \times (N-1)$ matrix model, 
and the expectation value $\bracket{\mathcal O}$ is defined as
\begin{equation}
 \bracket{O} = \frac{1}{Z}\int\rmd\Phi\,\mathcal O\,\e^{-N\tr V(\Phi)}.
\end{equation}
In the large-$N$ limit, the system of an $(N-1)\times (N-1)$ matrix 
is the same as the system of an $N\times N$ matrix. 
Hence, we can use the standard $N\times N$ matrix model 
to calculate these expectation values. 

The effective potential $V_\eff(x)$ defined above can be expressed 
in terms of connected diagrams. 
After some algebra, we obtain
\begin{equation}
 NV_\eff(x) = NV(x) - 2\bracket{\tr\log(x-\Phi)}_c
  - \frac{1}{2}\bracket{(2\tr\log(x-\Phi))^2}_c - \cdots.
\end{equation}
Here, the subscript ``$c$'' indicates the connected part. 
In the large-$N$ limit, the first and second terms are of order $N$ 
and third term is of order $N^0$. 
If we restrict ourselves to the leading order of $N$, 
the terms other than the first two can be neglected. 
Using the resolvent\footnote
{Here the branch of the square root is chosen so that 
$R(x)\sim 1/x$ as $x\rightarrow\infty$.} 
\begin{equation}\label{Resolvent}
 R(x) = \bracket{\frac{1}{N}\tr\frac{1}{x-\Phi}}
  = \frac{1}{2}\left(V'(x)-\sqrt{V^{\prime 2}(x) + p(x)}\right),
\end{equation}
the equation can be expressed as follows: 
\begin{subequations}
 \begin{align}
  NV_\eff(x) &= NV_\eff^{(0)}+V_\eff^{(1)}+\frac{1}{N}V_\eff^{(2)}+\cdots , \\
  V_\eff^{(0)} &= V(x) - \frac{1}{N}\bracket{2\tr\log(x-\Phi)} \\
  &= V(x) - 2\Repa\int^x\rmd x'R(x')
  = \Repa\int^x\rmd x'\sqrt{V^{\prime 2}(x')+p(x)}.
 \end{align}
\end{subequations}
The resolvent $R(x)$ has the cut on the real axis. 
If $x$ is on the cut, 
the effective potential becomes constant and 
the eigenvalue density $\rho(x)$ takes a nonzero value. 
Interpreting this in physical terms, 
we are considering the $N$-th eigenvalue $x$
and the other $N-1$ eigenvalues are those of $\Phi$ 
whose distribution is expressed by $\rho(x)$. 
In the cut, where the $N-1$ eigenvalues are distributed, 
the forces from them acting on the $N$-th eigenvalue 
cancel each other. 
The effective potential $V_\eff(x)$ 
has is at the minimum over the entire cut. 
{}From the standpoint of the original system of 
the $N\times N$ matrix including the $N$-th eigenvalue, 
it is natural that the eigenvalue density $\rho(x)$ should not  
change due to the $N$-th eigenvalue at the leading order of $N$; 
that is, the back reaction is at the subleading order. 
Hence, in the integration with respect to $x$ 
in \eqref{EffectivePotential}, most of 
the contribution comes from the region where $x$ is inside the cut. 
In the case of one cut, the integration over this region  
gives the partition function of the $N\times N$ matrix system. 
To extend this to type 0B string theory, 
we should consider the case of two cuts. 
In the case of two cuts, 
there are two minima of the effective potential. 
Because the potential under consideration is 
$Z_2$ symmetric, 
these two minima should be symmetric under $x\to -x$ 
and give the same contribution to the partition function. 
Hence, we can deal with these two regions together as 
``inside the cut.'' 
There is another nonzero contribution from the region 
where $x$ lies outside the cut. 
Because there is a maximum of the effective potential, 
we should take this into account. 
If the $N$-th eigenvalue $x$ lies outside the cut, 
the eigenvalue density of the $N\times N$ matrix system 
differs from that of the $(N-1) \times (N-1)$ matrix system. 
It can be identified as the ``instanton'' characterized 
by the configuration where $x$ is located at the local maximum 
of the effective potential. 
The resolvent is related to the disk amplitude of 
the Liouville field theory with the boundary condition 
corresponding to the FZZT-brane \cite{Fateev:2000ik}. 
The effect of the instanton comes from 
the (integrated) resolvent with some fixed value of 
the cosmological constant corresponding to the local maximum 
of the effective potential. 
It is related to the ZZ-brane \cite{Zamolodchikov:2001ah}. 

Now, we divide the integration region into two parts, 
namely, inside the cut and outside the cut, as 
\begin{align}
  Z &= Z_{N-1}\int_{\text{inside the cut}}\rmd x
   \bracket{\det(x-\Phi)^2}\e^{-NV(x)} \notag\\
   &\quad +Z_{N-1}\int_{\text{outside the cut}}
   \rmd x\bracket{\det(x-\Phi)^2}\e^{-NV(x)}.
\end{align}
The second term comes from the integration outside the cut, 
and is identified as the contribution of the instanton. 
So far, we have restricted ourselves to the $N$-th eigenvalue $x$ 
and identified the instanton as the configuration 
where $x$ lies outside the cut. 
However, there are $N$ eigenvalues and 
all other $N-1$ eigenvalues can be possibly outside the cut as well. 
Hence, there is an $n$-instanton sector; that is, $n$ eigenvalues 
lie outside the cut. 
The partition function can be expressed the sum of 
those in the $n$-instanton sector as
\begin{subequations}
 \begin{align}
 Z &= Z^{(\text{0-inst})}+Z^{(\text{1-inst})}+Z^{(\text{2-inst})}+\cdots , \\
 Z^{(\text{0-inst})} &= \int_{\text{inside the cut}}
 \prod_i\rmd\lambda_i\,
 \Delta^2(\lambda)\,\e^{-N\ssum{i}{}V(\lambda_i)}\notag\\
 &= Z^{(\text{0-inst})}_{N-1}\int_{\text{inside the cut}}\rmd x
 \bracket{\det(x-\Phi)^2}^{(\text{0-inst})}\e^{-NV(x)} , \\
 Z^{(\text{1-inst})} &= NZ^{(\text{0-inst})}_{N-1}
 \int_{\text{outside the cut}}\rmd x
 \bracket{\det(x-\Phi)^2}^{(\text{0-inst})}\e^{-NV(x)}.
 \label{1-inst Z}
 \end{align}
\end{subequations}
Here, the superscript ``($n$-inst)'' indicates the $n$-instanton sector. 
In the partition function  
$Z^{(\text{0-inst})}$ and the expectation value 
$\bracket{\mathcal O}^{(\text{0-inst})}$ in the 0-instanton sector, 
all eigenvalues are inside the cut. 
Hereafter we will omit the superscript ``(0-inst)'' in the 
expectation value.
The factor $N$ in front of the 1-instanton sector partition function 
\eqref{1-inst Z}
reflects the number of ways of specifying 
an eigenvalue that lies outside the cut. 
If we consider all $n$-instanton sectors, 
neglecting interaction between instantons, which is valid 
in the large-$N$ limit, 
the partition function can be expressed in terms of 
those of the 0-instanton and the 1-instanton sector
\begin{subequations}
 \begin{align}
  e^{F} = Z &= Z^{(\text{0-inst})}
  \left(1+\frac{Z^{(\text{1-inst})}}{Z^{(\text{0-inst})}}+\cdots\right)
  = \e^{F^{(\text{0-inst})}+\mu} , \\
  &\mu = \frac{Z^{(\text{1-inst})}}{Z^{(\text{0-inst})}}
  = N\frac{\int_{\text{outside the cut}}\rmd x
  \bracket{\det(x-\Phi)^2}\e^{-NV(x)}}
  {\int_{\text{inside the cut}}\rmd x
  \bracket{\det(x-\Phi)^2}\e^{-NV(x)}}. \label{ChemicalPotential}
 \end{align}
\end{subequations}
The additional term $\mu$ is the contribution from instantons. 
We regard this term as the ``chemical potential'' of the instanton. 
This effect corresponds to the ZZ-brane 
in the noncritical string theory.

\section{Effective potential and orthogonal polynomials}\label{sec:orth_poly}

In the previous section we have seen that the partition function 
has contributions which come from the multi-instanton sectors,  
that are characterized as configurations where 
some of the eigenvalues are located at the maximum 
of the effective potential. 
On the other hand, in order to compute the chemical potential 
of the instanton, 
which is nothing but the instanton effect in the free energy, 
it is sufficient to consider the effect of only one instanton. 
This amounts to considering $\bracket{\det(x-\Phi)^2}$ 
in the 0-instanton sector and performing the integration 
with respect to $x$ as shown in \eqref{ChemicalPotential}. 
However, the integration of $\bracket{\det(x-\Phi)^2}$ 
inside the cut generally contains divergent contributions 
in the subleading order of $N$ \cite{Hanada:2004im}. 
Our expectation here is that these divergences cancel the overall $N$ 
in \eqref{ChemicalPotential} to make the chemical potential finite. 
In order to confirm this, we have to retain the $N$-dependence 
in the computation. For this purpose, it is appropriate 
to use the method of orthogonal polynomials, because 
it is available for any $N$.  
  
We begin with definitions and properties of the orthogonal 
polynomials. 
The partition function of the one-matrix model can be expressed 
in terms of orthogonal polynomial $P_n(x)$ as 
\begin{equation}
  Z = \int\prod_i\rmd\lambda_i\,
 \det_{nn'}P_{n}(\lambda_{n'})\det_{mm'}P_m(\lambda_{m'})\,
 \e^{-N\ssum{i}{}V(\lambda_i)}.
\end{equation} 
Here, the orthogonal polynomial $P_n(x)=x^n+\mathcal O(x^{n-1})$ 
satisfies the orthogonality condition 
\begin{equation}
 (P_n(x),P_m(x)) = \int\rmd x\,P_n(x)P_m(x)\,\e^{-NV(x)} = h_n\delta_{nm}.
\end{equation}
Using this inner product of the orthogonal polynomials, 
the partition function can be expressed as 
\begin{equation}
 Z = N!\det_{nm}(P_n,P_m) = N!\,h_0^N\prod_{n=1}^{N-1} f_n^{N-n}.
  \label{PartitionFunction}
\end{equation}
Here, $f_n = h_n/h_{n-1}$. 
The orthogonal polynomials can be determined 
by recursion relations
\begin{subequations}\label{RecursionRelation}
 \begin{align}
  xP_n(x) &= X_{nm}P_m(x) = P_{n+1}(x) + s_n P_n(x) + r_n P_{n-1}(x), 
  \label{Recursion1}\\
  P_n'(x) &= \mathcal P_{nm}P_m(x) = \left[NV'(X)\right]_{nm}P_m(x).
  \label{Recursion2}
 \end{align}
\end{subequations}
It can be easily seen that $r_n = f_n$. 
Eliminating $r_n$ and $s_n$, we will obtain 
a differential equation that determines $P_n(x)$. 
However, we use these relations in a slightly different way. 
Because the free energy is expressed in terms of $r_n$'s, 
we should determine them. 
This can be done using \eqref{Recursion2}. 
Indeed, $r_n$ and $s_n$ can be determined by \eqref{Recursion2}, 
then using \eqref{Recursion1}, $P_n(x)$ can be expressed 
in terms of $r_n$ and $s_n$. 

In the large-$N$ limit, it is natural that 
any quantity $f_n$ with index $n$ 
can be approximated by a continuous function $f(\xi)$ 
with $\xi = n/N$. 
In fact, $f_n$ becomes continuous in the case of one cut. 
In order to do this, however, the values of $f_n$ and $f_{n+1}$ 
should become closer in the large-$N$ limit. 
In the case of two cuts, 
$f_n$ cannot be approximated by a continuous function, 
but if we consider $f_n$ with even $n$ or odd $n$ separately, 
they can be approximated by different continuous functions 
\cite{Demeterfi:1990gb}. 
We use $\widehat f_n$ to indicate 
that the index $n$ is even, and $\widetilde f_n$ for odd $n$. 
In the large-$N$ limit, 
they are approximated by different functions as 
\begin{equation}
 f_n = 
 \begin{cases}
  \widehat f_n = \widehat f (\xi) & (n:\text{even}) \\
  \widetilde f_n = \widetilde f (\xi) & (n:\text{odd})
 \end{cases}.
\end{equation}
In this limit, a summation over the index $n$ 
can be approximated by an integration. 
If we want to calculate up to the next-to-leading order 
in $1/N$, we should use the Euler-Mclaurin 
summation formula. 
For example, for even $N$ 
\begin{equation}
 \sum_{n=0}^N f_n =
 \frac{N}{2}\int_{0-\frac{1}{N}}^{1+\frac{1}{N}}\rmd\xi\,\widehat f(\xi)
 +\frac{N}{2}\int_0^1\rmd\xi\,\widetilde f(\xi).
\end{equation}
In this manner we treat corrections of the next-to-leading order 
in a systematic way. 

Now, we compute the chemical potential of the instanton 
\begin{equation}
 \mu
 = N\frac{\int_{\text{outside the cut}}\rmd x
 \bracket{\det(x-\Phi)^2}\e^{-NV(x)}}
 {\int_{\text{inside the cut}}\rmd x
 \bracket{\det(x-\Phi)^2}\e^{-NV(x)}}. \tag{\ref{ChemicalPotential}}
\end{equation}
Here, it is easy to show that the expectation value 
$\bracket{\det(x-\Phi)}$ can be identified with $P_n$ 
\begin{equation}
 P_n(x) = \bracket{\det(x-\Phi)}_n,
 \label{P_n} 
\end{equation}
where the subscript ``$n$'' again indicates a quantity 
for an $n\times n$ matrix system. 
Because the quantity under consideration is not 
the expectation value of the trace, but of the determinant,  
the large-$N$ factorization does not hold in this case. 
In fact, if we define $D_n(x)$ as 
\begin{equation}
 D_n(x) = \bracket{\det(x-\Phi)^2}_n ,
\end{equation}
it satisfies a recursion relation 
\begin{equation}
 D_n = P_n^2(x) + r_n D_{n-1}.
\end{equation}
Using this relation recursively, 
we obtain 
\begin{equation}
 D_N(x) = P_N^2(x) + r_N P_{N-1}^2(x) + \cdots 
  + r_N \cdots r_1 P_0^2(x). \label{formula4D_N}
\end{equation}
This formula enables us to evaluate the chemical potential. 
Substituting this formula into \eqref{ChemicalPotential}, 
the chemical potential can be expressed in terms of 
the orthogonal polynomials, which can be determined by 
\eqref{RecursionRelation}, and we will obtain 
the definite value of the chemical potential.

When we evaluate $\bracket{\det(x-\Phi)^2}$, 
the result will be different depending on whether $x$ is inside the cut or 
outside the cut. 
This difference can be described as follows. 
The orthogonal polynomial $P_n$ can be expressed 
as the expectation value in the 
$n\times n$ matrix system as in \eqref{P_n}, 
and thus if $x$ is inside the cut of this system, 
$P_n(x)$ is oscillating rapidly, otherwise 
$P_n(x)$ is monotonic as $x^n$. 
As $n$ increases, the oscillatory region becomes wider. 

When $x$ is inside the cut, 
we define $n_*(x)$ as the minimum value of $n$ 
such that $x$ is inside this oscillatory region of $P_n(x)$. 
The normalized orthogonal polynomials 
$\psi_n(x) = P_n(x)/\sqrt{h_n}$ 
take values of the same order for $n\geq n_*(x)$, 
while those for $n < n_*(x)$ $\psi_n(x)$ become small 
so that the contributions from the last $n_*(x)$ terms 
containing $P_n(x)$ ($n<n_*(x)$) in \eqref{formula4D_N} 
are damped exponentially.  
Hence, we can neglect the term of $P_n$ 
with $n< n_*(x)$ in \eqref{formula4D_N}. 
On the other hand, if $x$ is outside the cut, 
$P_n(x)$ for all $n$ are not oscillating but monotonic 
with respect to $x$ as $x^n$ and we cannot neglect the latter terms 
with $P_n(x)$ for $n< n_*(x)$. 
Thus, in the computation of \eqref{formula4D_N}, we should consider 
these two cases, namely, inside the cut and outside the cut, 
separately. 

First, we consider the case in which $x$ is outside the cut. 
In this case, the largest contribution  
to $\bracket{\det(x-\Phi)^2}$ comes from the first term in 
\eqref{formula4D_N} and contributions from latter terms 
become smaller like a geometric series. 
Using the ratio of the orthogonal polynomials 
\begin{equation}
 \e^{k_n} \equiv \frac{P_n(x)}{P_{n-1}(x)},
\end{equation}
$D_N = \bracket{\det(x-\Phi)^2}$ can be expressed as 
\begin{equation}
 D_N = P_N^2(x)
  \left[
   1 + r_N\e^{-2k_N} + r_N r_{N-1}\e^{-2k_N-2k_{N-1}}+\cdots
  \right].
\end{equation}
In the case of one cut, 
we can approximate this by a geometric series 
\begin{equation}
 D_N = \sum_{n=0}^\infty P_N^2(x)
  \left[r_N\,\e^{-2k_N}\right]^n 
  \left(1+{\cal O}\left(\frac{1}{N}\right)\right).
\end{equation}
In the case of two cuts, 
however, we should distinguish a quantity with even $n$ and odd $n$. 
Hence, we approximate this up to ${\cal O}(1/N)$ 
by a sum of two geometric series as  
\begin{subequations}
 \begin{align}
  D_N &\simeq \sum_{n=0}^{\infty} P_N^2
  \left(1 + \widehat r_N\e^{-2\widehat k_N}\right)
  \left[\widehat r_N \widetilde r_N 
  \e^{-2(\widehat k_N+\widetilde k_N)}\right]^n \\
  &= \frac{P_N^2(x)\left(1 + \widehat r_N\e^{-2\widehat k_N}\right)}
  {1-\widehat r_N \widetilde r_N 
  \exp\left[-2(\widehat k_N+\widetilde k_N)\right]}
  ,
 \end{align}
\end{subequations}
for even $N$, and 
\begin{subequations}
 \begin{align}
  D_N &\simeq \sum_{n=0}^{\infty} P_N^2
  \left(1 + \widetilde r_N\e^{-2\widetilde k_N}\right)
  \left[\widehat r_N \widetilde r_N 
  \e^{-2(\widehat k_N+\widetilde k_N)}\right]^n \\
  &= \frac{P_N^2(x)\left(1 + \widetilde r_N\e^{-2\widetilde k_N}\right)}
  {1-\widehat r_N \widetilde r_N 
  \exp\left[-2(\widehat k_N+\widetilde k_N)\right]}
  , 
 \end{align}
\end{subequations}
for odd $N$. 
In the case of two cuts, 
we obtain (see Appendix \ref{app:orth_poly})
\begin{subequations}\label{formula4k}
 \begin{align}
  k_n &= k_n^{(0)} + \frac{1}{N}k_n^{(1)} + \cdots , \\
  &\e^{\widehat k_n^{(0)}} = \frac{1}{2x}
  \left\{(x^2+\widehat r_n-\widetilde r_n)
  \pm\sqrt{(x^2-\widehat r_n-\widetilde r_n)^2
  -4\widehat r_n \widetilde r_n}\right\} , \label{khatdef}\\
  &\e^{\widetilde k_n^{(0)}} = \frac{1}{2x}
  \left\{(x^2-\widehat r_n+\widetilde r_n)
  \pm\sqrt{(x^2-\widehat r_n-\widetilde r_n)^2
  -4\widehat r_n \widetilde r_n}\right\} , \label{ktildedef}\\
  &\widehat k_n^{(1)} + \widetilde k_n^{(1)} 
  = -\frac{1}{2}\partial_\xi\log
  \left[(x^2-\widehat r_n-\widetilde r_n)^2
  -4\widehat r_n \widetilde r_n\right] ,
 \end{align}
\end{subequations}
where the double sign is understood to be $+$ for 
$x^2 > \widehat r_n+\widetilde r_n+2\sqrt{\widehat r_n \widetilde r_n}$ 
so that 
$e^{k_n^{(0)}}\sim x$ as $|x|\rightarrow \infty$, 
and $-$ for 
$x^2 < \widehat r_n+\widetilde r_n-2\sqrt{\widehat r_n \widetilde r_n}$ 
so that $e^{k_n^{(0)}}$ does not diverge at $x=0$. 
Using these relations, we obtain 
\begin{subequations}\label{OutsideCut}
 \begin{align}
   D_N &= \left\lvert
  \frac{\left(x^2\mp\widehat r_N \pm \widetilde r_N\right)
  \exp\left[\widehat k_N^{(0)}+\widetilde k_N^{(0)}\right]}
  {(x^2-\widehat r_N - \widetilde r_N)^2-4\widehat r_N \widetilde r_N}
  \right\rvert\,\e^{NW(x)} 
  \left(1+{\cal O}\left(\frac{1}{N}\right)\right), \\
  W(x) &= \int_0^1\rmd\xi
 \left(\widehat k^{(0)}(\xi) + \widetilde k^{(0)}(\xi)\right)
  = 2\int^x\rmd x'R(x) . \label{KandResolvent}
 \end{align}
\end{subequations}
Here, the sign ``$\pm$'' distinguishes the cases of even $N$  
and odd $N$. 
The relation between $W(x)$ and resolvent $R(x)$ can be checked 
with a concrete potential $V(x)$, or in the double scaling limit. 
We will examine this point later.

Next, we consider the case where $x$ is inside the cut. 
In this case, the contribution from $P_n(x)$ for $n < n_*(x)$ 
is exponentially suppressed. 
Hence, $D_N=\bracket{\det(x-\Phi)^2}$ can be approximated as 
\begin{equation}
 D_N = \sum_{n=n_*(x)}^N P_n^2(x) \prod_{m=0}^{N-n-1} r_{N-m} .
 \label{D_Nincut}
\end{equation}
In the large-$N$ limit, the summation over $n$ 
can be replaced by an integration. 
By definition, $x$ is in the 
oscillatory region of $P_n(x)$ with $n \geq n_*(x)$. 
Because its frequency is of the order $N$ as shown in 
Appendix \ref{app:orth_poly}, in order to calculate $P_n(x)^2$ 
in the large-$N$ limit it is sufficient to take the average 
of this oscillation, which amounts to dividing the amplitude 
of $P_n(x)^2$ by 2. 
In the case of two cuts, 
generally, we should distinguish a quantity for even $n$ 
and a quantity for odd $n$. 
However, after we average this oscillation, 
$\psi_n(x)=P_n(x)/\sqrt{h_n}$ can be 
approximated by a continuous function of $\xi=n/N$ 
whether $n$ is even or odd. 
Finally we should take the extra factor 2 into account. 
This factor comes from a relative normalization 
between the $P_n(x)$ inside the cut and $P_n(x)$ outside the cut 
as in the continuity formula in the usual WKB approximation 
\cite{Hanada:2004im}. 
Using the asymptotic formula for $\psi_n(x)$ inside the cut 
given in Appendix \ref{app:orth_poly}, \eqref{D_Nincut} takes a value 
up to ${\cal O}(1/N)$ in the exponent
\begin{subequations}\label{InsideCut}
 \begin{align}
  D_N &= \sum_{n=n_*}^N h_N \psi_n^2(x) \prod_{m=0}^{N-n-1} r_{N-m}\\
  &= 2\e^{N\Repa W(x)}
  \sqrt{r_N}N\int_{\xi_*}^{1}\rmd\xi\frac{x}
  {\sqrt{4\widehat r(\xi) \widetilde r(\xi)
  -\left(x^2-\widehat r(\xi) -\widetilde r(\xi)\right)^2}}\\
  &= 2\pi N\rho(x)\sqrt{r_N}\,\e^{N(V(x)+W_0)}, 
 \end{align}
\end{subequations}
where $\xi_*=n_*/N$ and $W(x)$ is again related to the resolvent 
as in \eqref{KandResolvent}. 
Because the real part of the resolvent becomes $V'(x)$ 
inside the cut, 
$\Repa W(x) = V(x)+W_0$,  
where $W_0$ is a constant that determines 
the origin of $V_\eff^{(0)}(x)$. 
Note that because $\Repa W(x)$ does not depend on $\xi=n/N$ 
for $n\geq n_*$, as shown in Appendix \ref{app:orth_poly}, 
we can put it outside the integration with respect to $\xi$. 
We have also used the fact that 
the integration in the second line can be identified 
with the eigenvalue density $\rho(x)$, 
which is shown in Appendix \ref{app:VeffandR}. 

Now, we are ready to take the ratio of the partition functions 
inside and outside the cut, and obtain 
the chemical potential of the instanton. 
This can be expressed as 
\begin{equation}
 \mu
 = N\frac{\int_{\text{outside the cut}}\rmd x
 \bracket{\det(x-\Phi)^2}\e^{-NV(x)}}
 {\int_{\text{inside the cut}}\rmd x
 \bracket{\det(x-\Phi)^2}\e^{-NV(x)}} . \tag{\ref{ChemicalPotential}}
\end{equation}
Most of the contribution from outside the cut comes 
from the saddle point of the effective potential 
in the large-$N$ limit. 
In our case, there is a maximum of the effective potential 
at $x=0$. This maximum always exists when we consider the case 
where there are two cuts that are symmetric under $x\to -x$. 
We consider only the instanton that is located at this maximum. 
Using the method of steepest descent, 
we obtain the chemical potential with an overall factor 
of ${\cal O}(N^0)$ as  
\begin{equation}
 \mu = i\sqrt{\frac{\pi}{N(R'(0)-\frac{1}{2}V''(0))}}\frac
 {\left\lvert\widehat r_N + \widetilde r_N 
   + \left\lvert\widehat r_N - \widetilde r_N \right\rvert\right\rvert}
 {4\pi\sqrt{\widehat r_N}
 \left\lvert\widehat r_N - \widetilde r_N \right\rvert}
 \exp\left[{N\int_a^0\rmd x\,\left\{2R(x)-V'(x)\right\}}\right],
\end{equation} 
where $a$ corresponds to $W_0$ and can be chosen as any point 
in the left one of the two cuts. 
As mentioned above, it indeed determines the origin of 
$V_\eff^{(0)}(x)$. 
If there is another maximum or minimum of the effective potential, 
we should take it into account in the contribution from outside the cut. 
Under the double scaling limit, 
however, there can be no contribution from such a saddle point. 
Hence, it is sufficient to consider the instanton 
at the maximum of $x=0$. 
We will elucidate this point further in the next section.

\section{Universality of chemical potential}
\label{sec:universality}

In this section, we consider the double scaling limit 
in the case of two cuts. 
This is obtained by taking the limits 
$N\to\infty$ and $g\to g_c$ with a certain combination of them fixed. 
Here, $g$ is a parameter of the potential $V(x)$ and 
$g_c$ is its critical value. 
The critical point around which the type 0B noncritical string 
is described is one of the Gross-Witten phase transitions.  
It can be found as the critical point of the one-matrix model 
where two cuts, which are the supports of the eigenvalue distribution, 
become closer and are merged. 
If this critical point is exceeded, 
we have a one-cut eigenvalue distribution. 
Hence, this critical point distinguishes two phases, 
namely, the one-cut phase and the two-cut phase. 

Before taking up the double scaling limit, 
we consider the behavior of the matrix model 
near the critical point in the two-cut case. 
To compute quantities with the method of orthogonal polynomials, 
it is necessary to evaluate the value of $r_n$. 
This can be done by using \eqref{Recursion2}. 
Because $P_n(x)$ is monic, $\mathcal P_{n,n-1}=n$. 
Picking up the coefficient of $P_{n-1}$ in \eqref{Recursion2}, 
we obtain at the leading order of $1/N$
\begin{equation}
 g\xi = F(\widehat r,\widetilde r) = F(\widetilde r,\widehat r). 
 \label{FandXi}
\end{equation}
Here, $F(x,y)$ originates from $\left[gV'(X)\right]_{n,n-1}$. 
In fact, there are two relations derived from \eqref{Recursion2}, 
namely, one for even $n$ and one for odd $n$. 
If we define $F(\widehat r,\widetilde r)$ for even $n$, 
the same variable for odd $n$ can be obtained by 
interchanging $\widehat r$ and $\widetilde r$, 
which is nothing but $F(\widetilde r,\widehat r)$. 
At the critical point, the two cuts merge and 
$\widehat r$ and $\widetilde r$ take the same value $r_c$. 
This means that $\widehat r = \widetilde r = r_c$ for $n=N$ 
at the critical point. 
Hence, we obtain $g_c=F(r_c,r_c)$. 
Expanding \eqref{FandXi} around $r_c$, 
it can be expressed as 
\begin{align}
 F(\widehat r,\widetilde r)
 &= g_c - \widehat A (r_c-\widehat r) 
 - \widetilde A (r_c-\widetilde r) \notag\\
 &\quad - \widehat B (r_c -\widehat r)^2 - \widetilde B (r_c - \widetilde r)^2
 - C (r_c - \widehat r)(r_c - \widetilde r) .\label{DefF}
\end{align}
It is necessary for at least one of following conditions to be 
satisfied so that 
$F(\widehat r, \widetilde r)= F(\widetilde r,\widehat r)$ 
holds:

\begin{description}\setlength{\itemindent}{0pt}
 \item[Case 1:] $\widehat r = \widetilde r$
 \item[Case 2:] $\widehat A = \widetilde A$ and
	    $(r_c-\widehat r)+(r_c -\widetilde r)=0$
 \item[Case 3:] $\widehat A = \widetilde A$ and
	    $\widehat B = \widetilde B$.
\end{description}
The condition that is relevant to the two-cut case is case 2. 
In this case, if $g\leq g_c$, a solution of \eqref{FandXi} 
with real $r$'s exists that can be rewritten as  
\begin{subequations}\label{universalR}
 \begin{align}
  \widehat r &= r_c - \alpha\sqrt{g_c-g\xi}, \\
  \widetilde r &= r_c + \alpha\sqrt{g_c-g\xi} .
 \end{align}
\end{subequations}

For $g > g_c$, we cannot take this condition. 
In this case, the condition $\widehat r = \widetilde r$ 
will be satisfied, which is the case of one cut. 
Using the definition in \eqref{universalR}, 
a solution of \eqref{FandXi} in this case 
can be expressed as (see Appendix~\ref{app:relation})
\begin{equation}
  r = r_c - \frac{\alpha^2}{4r_c}(g_c-g\xi) .
\end{equation}

First, we evaluate the free energy 
without the contribution from the instanton. 
In the Gross-Witten phase transition 
comprising the third order, 
the third derivative of the free energy in general 
has a discontinuity, which is universal in the sense 
that it is not affected by details of the potential. 
Thus, in order to extract the universal part of the free energy, 
we will compare the free energy of each phase|one cut 
and two cuts|and pick up a discontinuity of (the third derivative of) 
the free energy, and which is the only contribution 
to the universal part. 
For the free energy in the two-cut phase, 
from \eqref{PartitionFunction} we obtain 
\begin{subequations}\label{2cutFreeEnergy}
 \begin{align}
   F &= \frac{N^2}{2}\int_0^1\rmd\xi\,(1-\xi)
 \left[\log \widehat r(\xi) + \log \widetilde r(\xi) \right]\\
 &\simeq -\frac{N^2}{2}\int_0^1\rmd\xi\,(1-\xi)
 \left(\frac{\alpha^2}{r_c}\right)^2(g_c-g\xi) \\
 &\simeq -\frac{N^2}{12}\left(\frac{\alpha}{gr_c}\right)^2(g_c-g)^3 , 
  \label{2cutFreeEnergyWithG}
 \end{align}
\end{subequations}
and for the one-cut phase, 
\begin{subequations}\label{1cutFreeEnergy}
 \begin{align}
  F &= N^2\int_0^1\rmd\xi\,(1-\xi) \log r(\xi)\\
  &\simeq -N^2\int_{\xi_0}^1\rmd\xi\,(1-\xi)
  \left(\frac{\alpha^2}{4r_c}\right)^2(g_c-g\xi) \notag\\
  &\quad -\frac{N^2}{2}\int_0^{\xi_0} \rmd\xi\,(1-\xi)
  \left(\frac{\alpha^2}{r_c}\right)^2(g_c-g\xi) \\
  &\simeq -\frac{N^2}{24}\left(\frac{\alpha}{gr_c}\right)^2(g_c-g)^3 . 
  \label{1cutFreeEnergyWithG}
 \end{align}
\end{subequations}
Here, we have omitted the terms that do not contribute to 
the discontinuity of the free energy. 
In computation of the free energy in the one-cut phase, 
we have noticed the following fact:
in the one-cut phase, in general there is only one cut 
constructed from $N$ eigenvalues. 
However, if there were $n<n_0=Ng_c/g$ eigenvalues, 
the cut would split into two. 
Hence, for $\xi < \xi_0 \equiv n_0/N$, 
we should take the two-cut solution \eqref{universalR}, 
even in the one-cut phase. 
Comparing \eqref{2cutFreeEnergyWithG} and 
\eqref{1cutFreeEnergyWithG}, 
we can see that the third derivative of the free energy 
indeed has a discontinuity. 

Second, we consider the contribution from the instanton 
in a background with a fixed number of instantons. 
In the one-instanton background, 
an instanton is located at the top of the effective potential. 
The contribution to the free energy from this instanton 
can be obtained by the height of the potential barrier. 
The effective potential can be obtained from the resolvent 
that is expressed near the critical point as 
\begin{subequations}
 \begin{align}
  R(x) &= \frac{1}{2}\partial_x\int_0^1\rmd\xi
  \left[\widehat k^{(0)}(\xi) + \widetilde k^{(0)}(\xi)\right]
  \simeq C'x\sqrt{a^2-x^2}, \\
  & C' = \frac{\sqrt{r_c}}{\alpha^2 g}, \qquad\qquad
  a^2 = \frac{\alpha^2}{r_c}(g_c-g) ,
 \end{align}
\end{subequations} 
where we have dropped the contribution from 
the potential $V(x)$, which is nonuniversal. 
Then, the height of the potential barrier is 
\begin{equation}
 2N\int_a^0\rmd x R(x) = -\frac{2}{3}
 N\frac{\alpha}{gr_c}\left(g_c-g\right)^\frac32 . 
 \label{1instFreeEnergy}
\end{equation}
This is the contribution to the free energy from the instanton 
in the one-instanton background. 
In the $n$-instanton background, the contribution 
of the instantons at the leading order 
is multiplied by $n$. 

Third, we evaluate the chemical potential of the instanton. 
{}From \eqref{ChemicalPotential}, \eqref{universalR}, and 
\eqref{1instFreeEnergy}, we obtain 
\begin{equation}
 \mu = \frac{i}{4\sqrt{\pi N}}
 \frac{\sqrt{gr_c}}
 {\sqrt\alpha(g_c-g)^\frac34}
 \exp\left[-\frac{2}{3}
 N\frac{\alpha}{gr_c}\left(g_c-g\right)^\frac32\right] .
\end{equation}

Now, we are ready to consider the double scaling limit. 
Two limits, $N\to\infty$ and $g\to g_c$, are related as 
\begin{align}
  N &= a^{-3} & 
 \left(\frac{\alpha}{gr_c}\right)^\frac23(g_c-g) &= a^2 t &
 a&\to 0 .
\end{align}
There is an ambiguity in the definition of $t$. 
Here, we define $t$ so that  
the discontinuity of the free energy is described as  
$F''(t)=-|t|/8$. 
Using this definition, the free energy in our calculation and 
that in the string equation coincide. 
The string equation in this definition takes the form of 
the Painlev\'e II equation,
\begin{equation}
  th = h^3 - 2\partial_t^2 h . \label{PainleveII}
\end{equation} 
Here, $h(t)$ is related to the free energy as 
\begin{align}
  \partial_t^2 F &= - \frac{1}{2} h^2 + f & 
  t &= h^2 -4f .
\end{align}
The solution of \eqref{PainleveII} in the two-cut phase 
is given by 
\begin{equation}
 h(t) = \sqrt t + \cdots
\end{equation}
in perturbative expansion. 
The solution in the one-cut phase is $h(t)=0$. 
Thus, the discontinuity of the free energy is consistent 
with ours. 

In this double scaling limit, 
the chemical potential of the instanton becomes 
\begin{equation}
 \mu =  \frac{i}{4\sqrt\pi t^\frac34}
 \exp\left[-\frac{2}{3}t^\frac32\right] . 
 \label{ChemicalPotentialInDoubleScalingLimit}
\end{equation}
As shown above, the value does not depend on the details of the potential 
and hence it is a universal quantity. 
Moreover, it agrees with the nonperturbative effect 
derived from the string equation. This can be seen as follows.
If there are two solutions of the string equation 
with the same asymptotic expansion for large $t$
\begin{equation} 
h(t) = \sqrt t -\frac{1}{4}t^{-\frac{5}{2}}+ \cdots,
\label{asymptotic}
\end{equation} 
then the difference between them due to the nonperturbative effect 
is given by 
\begin{equation}
 h_{\text{inst}}(t) = \Const\times t^{-\frac{1}{4}}
  \e^{-\frac{2}{3}t^{\frac32}} .
\end{equation}
{}From this, the nonperturbative effect of the free energy 
can be obtained
as
\begin{equation}
 F_{\text{inst}} = \Const\times t^{-\frac34}\e^{-\frac{2}{3}t^\frac32} ,
\end{equation}
which is in accord with our chemical potential of the instanton. 
This justifies our identification of the instanton in the 
matrix model as the nonperturbative effect of string theory.
However, the overall normalization factor cannot be determined from 
the string equation itself. 
We have determined it from direct computations
in the matrix model and have shown that it is universal. 

\section{Conclusions}\label{sec:conclusion}

In this paper, we have investigated in full detail the 
nonperturbative effect in type 0B string theory, 
which is defined by taking the double scaling limit 
of the one-matrix model 
with a two-cut eigenvalue distribution.  
We have computed the contribution from the instanton 
to the free energy directly in the matrix model. 
In the double scaling limit, 
the chemical potential of the instanton does not depend on 
the details of the potential, 
and is a universal quantity. 
It takes exactly the same form as the nonperturbative effect 
derived from the string equation. 
Moreover, by computation via the matrix model keeping $N$ finite, 
we have fixed the overall factor of the chemical potential, 
which cannot be determined by the string equation itself. 
In \cite{Hanada:2004im}, it is shown that 
in (bosonic) $c=0$ noncritical string theory, only the closed string, 
or the string equation, does not describe the 
nonperturbative effect completely, 
and that the matrix model is more fundamental, 
capturing the nonperturbative effect correctly. 
Here, we have found that this is also the case with 
type 0B, or $\hat{c}=0$, noncritical string theory. 

It is worth noting a crucial difference 
between $c=0$ and $\hat{c}=0$ string theory. 
In the case of $c=0$ string theory, 
the potential of the matrix model is unbounded from below.  
Therefore, the vacuum with one cut is unstable and 
the imaginary part of the free energy 
obtained in \cite{Hanada:2004im} reflects this instability. 
On the other hand, in the case of $\hat{c}=0$ string theory 
defined by a matrix model with a two-cut eigenvalue distribution,  
the potential is a double-well type and bounded from below. 
Thus, the matrix integration in the definition 
of the partition function or the free energy as in \eqref{Z} 
is well defined and gives a real number. 
In fact, in the double scaling limit, 
the integration outside the cut in \eqref{ChemicalPotential} becomes
\begin{equation}
  \frac{\sqrt{r_c}}{2}\int_{-\sqrt t}^0\rmd\zeta
  \frac{\sqrt t}{t-\zeta^2}\,
  \e^{-\frac{2}{3}\left(t-\zeta^2\right)^\frac{3}{2}}. 
  \label{IntegralInDoubleScalingLimit}
\end{equation}
Therefore, the contribution from the instanton to the free energy 
is given by an integration of a real function, but it diverges.
This divergence can be attributed to the large-$N$ limit, 
where some eigenvalues are at the edge of the cut 
$\zeta=-\sqrt{t}$. 
Moreover, in the double scaling limit, 
some of the eigenvalues are pushed out of the cut 
$\zeta\ge -\sqrt{t}$, so that the edge of the eigenvalue distribution 
is smeared. 
In this sense, the boundary at $\zeta=-\sqrt{t}$ 
between the outside and inside of the cut becomes subtle, 
and it becomes somewhat artificial 
in the double scaling limit to divide the integration region 
into these two regions. Because this divergence originates from 
only a part of the eigenvalues that spread into the outside 
of the cut, it is less divergent compared to the integration 
over the inside of the cut given as the denominator 
in \eqref{ChemicalPotential}, which is of order $N$ as shown 
in \eqref{InsideCut}. Therefore, it may be possible to 
change the definition of ``inside the cut'' slightly 
and then renormalize the above divergence into the contribution 
in the denominator in \eqref{InsideCut}. In that case, 
it is important to confirm that the chemical potential 
is still universal irrespective of a slight change 
in the definition of the inside of the cut. 
However, at least as long as the string coupling constant $g_s$ 
is sufficiently small,\footnote
{$g_s$ can be restored on dimensional grounds as 
$t^{3/2}\rightarrow t^{3/2}/g_s$.} 
we can take advantage of the saddle point method to evaluate 
the above integration. Thus, the above integration is dominated 
by the saddle point $\zeta\sim 0$ and if we can choose the contour 
of $\zeta$ as the whole imaginary axis, 
the integration becomes finite and gives an imaginary number 
as in \eqref{ChemicalPotentialInDoubleScalingLimit}. 
The chemical potential obtained in this way is meaningful 
at least for small $g_s$. 
In fact, the instanton in the matrix model we have considered 
corresponds to the D-brane known as a ZZ-brane 
\cite{Zamolodchikov:2001ah}. This can be checked 
by noting that it gives an open boundary to the worldsheet, 
or more quantitatively, by comparing the disk amplitude 
in the fixed instanton background to that in the 
ZZ-brane background computed in the super Liouville theory 
as done in \cite{Hanada:2004im}. 
In type 0B string theory, the ZZ-brane is unstable. 
The chemical potential that we have computed as above is 
a purely imaginary number. 
It can be considered to reflect the fact that the D-brane 
under consideration is unstable. 
Although $\hat{c}=0$ string theory does not have the time direction 
and the instanton does not decay, 
if we consider the additional time direction associated with 
the energy of the system, the instanton will decay due to its instability. 
The chemical potential of the imaginary number 
can be considered to show this instability. 
In this sense, the chemical potential is analogous with 
the energy of a statistical system, where there is 
no time direction, but the energy gives the probability 
that the system is realized in the ensemble.  
Likewise, because the instanton corresponds to the ZZ-brane 
in $\hat{c}=0$ noncritical string theory, 
the chemical potential is related to the 
decay rate of the ZZ-brane. Note that it is not necessary 
for our result to agree with the computations 
in \cite{Douglas:2003up} of the decay rate 
of the ZZ-brane (D-particle) in $\hat{c}=1$ noncritical 
string theory, because the time in the target space 
is {\it a priori} not the same as the extra time direction 
associated with the free energy mentioned above. 

The most important problem remaining unsolved 
is to identify the boundary conditions of the string equation. 
Note that the matrix model from which the string equation 
is derived has no ambiguity. Therefore, it is natural 
to expect that the matrix model specifies some boundary 
conditions, by which we can determine the nonperturbative effect 
precisely. Because the string equation is a differential equation 
of the second order, there are two boundary conditions to be 
specified. It can be readily seen that the asymptotic behavior 
\eqref{asymptotic} needs fine-tuning of one parameter, 
and therefore there is one boundary condition remaining. 
It is not clear in general whether the contribution 
from the instanton to the free energy, or the nonperturbative 
effect in the free energy, has something to do with this 
boundary condition. In particular, if $g_s$ is sufficiently small, 
the nonperturbative effect is exponentially small 
and it is impossible in general to determine it as an ambiguity 
to be added to a perturbative series of the free energy, 
which is an asymptotic expansion. 
However, if we concentrate on 
the process where each term of a perturbative series is completely zero, 
we can identify the nonperturbative effect unambiguously 
like the tunneling effect. The chemical potential we have computed 
can be regarded as such an example. That is, it is purely imaginary 
and the free energy cannot have an imaginary part perturbatively. 
We therefore conclude that by defining the contour 
of the integration with respect to the eigenvalue 
as the imaginary axis, we can compute the chemical potential 
of the instanton, and this will give a boundary condition 
for the imaginary part of a complexified string equation. 
However, we have not yet fixed a boundary condition 
for the real part of the string equation. 
Note here that in order to identify the nonperturbative ambiguity, 
we should specify a choice of the contour in general as 
mentioned in \cite{DiFrancesco:1993nw}. 
 
Finally, it would be interesting to apply the computations 
employed in this paper to other matrix models; for example, 
the two-matrix model. This is left for future studies.

\subsubsection*{Acknowledgements:}

The authors would like to thank 
M.~Hanada, 
N.~Ishibashi, and T.~Matsuo 
for their fruitful discussions.
This work is supported 
in part by the Grant-in-Aid for Scientific Research (14540254) 
and the Grant-in-Aid for the 21st Century COE 
``Center for Diversity and Universality in Physics'' 
from the Ministry of Education, 
Culture, Sports, Science and Technology (MEXT) of Japan.
The work of T.K. is supported in part by 
the Special Postdoctoral Researchers Program.

\appendix

\section{Orthogonal polynomials}\label{app:orth_poly}

In this section, 
we derive the asymptotic behavior of the orthogonal polynomials 
both inside and outside the cut. 
This can be deduced using \eqref{RecursionRelation}. 
{}From \eqref{Recursion1}, we write the orthogonal polynomials 
in terms of $r_n$. 
Using the ratio of the orthogonal polynomials 
$e^{k_n}=P_n(x)/P_{n-1}(x)$, 
\eqref{Recursion1} can be expressed as 
\begin{equation}
 x = \e^{k_{n+1}} + r_n \e^{-k_n}. \label{eq4k}
\end{equation}
Here we have used the fact that $s_n$ is identically zero due to 
the symmetry with respect to $x\to -x$. 
In the case of two cuts, 
we should distinguish quantities for even $n$ 
and odd $n$ as follows: 
\begin{align}
  k_n &= 
 \begin{cases}
  \widehat k_n 
  & (\text{for even $n$}) \\
  \widetilde k_n 
  & (\text{for odd $n$})
 \end{cases},&
 r_n &= 
 \begin{cases}
  \widehat r_n 
  & (\text{for even $n$}) \\
  \widetilde r_n 
  & (\text{for odd $n$})
 \end{cases}.
\end{align}
{}From \eqref{eq4k} we obtain two relations, 
one for even $n$ and one for odd $n$: 
\begin{subequations}\label{eq4keo}
 \begin{align}
  x = \e^{\widetilde k_{n+1}} + \widehat r_n \e^{-\widehat k_n} &
  &(\text{for even $n$}), \label{eq4ke}\\
  x = \e^{\widehat k_{n+1}} + \widetilde r_n \e^{-\widetilde k_n} &
  &(\text{for odd $n$}) \label{eq4ko}.
 \end{align}
\end{subequations}
In the large-$N$ limit, $\widehat k_n$, 
$\widetilde k_n$, $\widehat r_n$, and $\widetilde r_n$ 
can be expanded as 
\begin{subequations}\label{largeNexpand}
 \begin{align}
  f_{n+1} &= f(\frac{n}{N}) 
  + \frac{1}{N}\partial_\xi f(\frac{n}{N}) + \cdots,
  \label{largeNtaylor}\\
  f_n &= f_n^{(0)} + \frac{1}{N} f_n^{(1)} + \cdots .
 \end{align}
\end{subequations}
Here, $f_n$ represents $\widehat k_n$, 
$\widetilde k_n$, $\widehat r_n$, or $\widetilde r_n$ 
and $f(\xi)$ is a continuous function of $\xi=n/N$ 
corresponding to $f_n$ in the large-$N$ limit. 
We can see that $\widehat r_n^{(1)}$ and $\widetilde r_n^{(1)}$ 
are identically zero from \eqref{Recursion2}. 
{}From now on, we will omit the superscript ``$(0)$'' 
for $\widehat r_n$ and $\widetilde r_n$. 
Substituting \eqref{largeNexpand} into \eqref{eq4keo}, 
we obtain 
\begin{subequations}\label{eq4kleading}
 \begin{align}
  x &= \e^{\widetilde k_{n}^{(0)}} 
  + \widehat r_n \e^{-\widehat k_n^{(0)}} &
  &(\text{for even $n$}),\\
  x &= \e^{\widehat k_{n}^{(0)}} 
  + \widetilde r_n \e^{-\widetilde k_n^{(0)}} &
  &(\text{for odd $n$}),
 \end{align}
\end{subequations}
at the leading order, and 
\begin{subequations}\label{eq4knextleading}
 \begin{align}
  0 &= \left(\widetilde k_n^{(1)}+\partial_\xi \widetilde k_n^{(0)}\right)
  \e^{\widetilde k_n^{(0)}} 
  - \widehat r_n \widehat k_n^{(1)}\e^{- \widehat k_n^{(0)}} &
  &(\text{for even $n$ }),\\
  0 &= \left(\widehat k_n^{(1)}+\partial_\xi \widehat k_n^{(0)}\right)
  \e^{\widehat k_n^{(0)}}
  - \widetilde r_n \widetilde k_n^{(1)}\e^{- \widetilde k_n^{(0)}} &
  &(\text{for odd $n$ }),
 \end{align}
\end{subequations}
at the next-to-leading order. 
{}From \eqref{eq4kleading}, 
we obtain 
\begin{subequations}\label{kleading}
 \begin{align}
  \e^{\widehat k_n^{(0)}} &= \frac{1}{2x}
  \left\{(x^2+\widehat r_n-\widetilde r_n)
  \pm\sqrt{(x^2-\widehat r_n-\widetilde r_n)^2
  -4\widehat r_n \widetilde r_n}\right\}, \\
  \e^{\widetilde k_n^{(0)}} &= \frac{1}{2x}
  \left\{(x^2-\widehat r_n+\widetilde r_n)
  \pm\sqrt{(x^2-\widehat r_n-\widetilde r_n)^2
  -4\widehat r_n \widetilde r_n}\right\},
 \end{align}
\end{subequations}
where the sign of $\pm$ is $+$ for 
$x^2 > \widehat r_n+\widetilde r_n+2\sqrt{\widehat r_n \widetilde r_n}$ 
so that 
$e^{k_n^{(0)}}\sim x$ as $|x|\rightarrow \infty$, 
and $-$ for 
$x^2 < \widehat r_n+\widetilde r_n-2\sqrt{\widehat r_n \widetilde r_n}$ 
not to diverge at $x=0$. 
Moreover, from \eqref{eq4knextleading}, 
\begin{equation}
 \widehat k_n^{(1)} + \widetilde k_n^{(1)} 
 = -\frac{1}{2}\partial_\xi\log
 \left[(x^2-\widehat r_n-\widetilde r_n)^2
 -4\widehat r_n \widetilde r_n\right] .\label{knextleading}
\end{equation}
By using these $k$s, the orthogonal polynomials can thus be expressed as 
\begin{equation}
 P_n(x) = \exp\left[\sum_{m=1}^n k_n\right].
\end{equation}
Because we need to perform computations up to 
the next-to-leading order of $1/N$, 
we use the Euler-Mclaurin summation formula to obtain 
\begin{subequations}\label{PEularMclaurin}
 \begin{align}
  P_n(x) &= 
  \exp\left[
  \frac{N}{2}\int_{0+\frac{1}{N}}^{1+\frac{1}{N}}\rmd\xi\,\widehat k(\xi)
  +\frac{N}{2}\int_0^1\rmd\xi\,\widetilde k(\xi)
  \right] &
  &(\text{for even $n$}), \\
  P_n(x) &= 
  \exp\left[
  \frac{N}{2}\int_{0+\frac{1}{N}}^1\rmd\xi\,\widehat k(\xi)
  +\frac{N}{2}\int_0^{1+\frac{1}{N}}\rmd\xi\,\widetilde k(\xi)
  \right] &
  &(\text{for odd $n$}).
 \end{align}
\end{subequations} 
Substituting \eqref{kleading} and \eqref{knextleading} into  
\eqref{PEularMclaurin}, 
we find 
\begin{subequations}
 \begin{align}
  P_n(x) &= \left(\frac{x^2}
  {(x^2-\widehat r_n - \widetilde r_n)^2
  -4\widehat r_n \widetilde r_n}\right)^\frac14
  \exp\left[\frac{N}{2}W_0\left(x,\frac{n}{N}\right)
  +\frac{1}{2}k_n^{(0)}
  \right] , \\
  W_0(x,\xi) &= \int_0^\xi\rmd\xi'
  \left(\widehat k^{(0)}(\xi') + \widetilde k^{(0)}(\xi')\right) .
  \label{P_nout}  
 \end{align}
\end{subequations}
Here, $k_n^{(0)}$ is $\widehat k_n^{(0)}$ for even $n$ 
and $\widetilde k_n^{(0)}$ for odd $n$. 

If $x$ is in the oscillatory region of $P_n(x)$, where 
$(x^2-\widehat r_n - \widetilde r_n)^2-4\widehat r_n \widetilde r_n<0$, 
$k_n$ is no longer real, but $P_n(x)$ should be real. 
We note here that $P_n(x)$ is a solution of a set of 
linear differential equations \eqref{RecursionRelation} 
and we have obtained its solution in a classically allowed region, 
\eqref{P_nout} via the WKB approximation. Therefore, in order to 
find $P_n$ in the oscillatory region which is an analog of 
a forbidden region, we invoke the continuity condition 
\cite{Hanada:2004im}. 
That is, it is sufficient to take a linear combination of 
two complex solutions of \eqref{Recursion1} 
to obtain the real solution, 
which is nothing but the real part of the naive expression 
in \eqref{P_nout} with an appropriate phase factor. 
We thus obtain $P_n(x)$ in the oscillatory region 
as 
\begin{align}
 P_n(x) &= 2\left(\frac{x^2r_n}
 {4\widehat r_n \widetilde r_n
 -(x^2-\widehat r_n - \widetilde r_n)^2}\right)^\frac14 \notag\\
 &\quad\times\exp\left[\frac{N}{2}\Repa W_0\left(x,\frac{n}{N}\right)
 \right]
 \sin\left[\Impa\left(\frac{N}{2}W_0(x,\frac{n}{N})
 +\frac{1}{2}k_n^{(0)}\right) + \theta
 \right] . \label{P_nin}
\end{align}
Here, we have used 
\begin{align}
 \Repa\widehat k_n^{(0)} &= \frac{1}{2}\log\widehat r_n, &
 \Repa\widetilde k_n^{(0)} &= \frac{1}{2}\log\widetilde r_n, 
 \label{Rek}
\end{align}
in the oscillatory region. 
The additional factor 2 comes from a relative normalization 
to the solution for the allowed region 
$(x^2-\widehat r_n - \widetilde r_n)^2-4\widehat r_n \widetilde r_n>0$ 
in the continuity condition, and $\theta$ is a constant phase factor. 
The quartic root is defined as real and positive, 
and the overall sign is included in $\theta$. 

If we consider the normalized orthogonal polynomials 
$\psi_n(x)=P_n(x)/\sqrt{h_n}$, 
This can be expressed as 
\begin{subequations}
 \begin{align}
  \psi_n(x) &= 2\left(\frac{x^2r_N}
  {4\widehat r_n \widetilde r_n
  -(x^2-\widehat r_n - \widetilde r_n)^2}\right)^\frac14 \notag\\
  &\quad\times\exp\left[\frac{N}{2}\Repa W_r\left(x,\frac{n}{N}\right)
  \right]
  \sin\left[\Impa\left(\frac{N}{2}W_0\left(x,\frac{n}{N}\right)
  +\frac{1}{2}k_n^{(0)}\right) + \theta
  \right], \\
 W_r\Bigl(x,&\frac{n}{N}\Bigr) = W_0\Bigl(x,\frac{n}{N}\Bigr)
  -\frac{1}{2}\int_0^{\frac{n}{N}}\rmd\xi'
  \left(\log\widehat r(\xi')+\log\widetilde r(\xi')\right).
 \end{align}
\end{subequations}
For $n\geq n_*(x)$, where $n_*(x)$ is defined by 
$(x^2-\widehat r_{n_*} - \widetilde r_{n_*})^2-4\widehat r_{n_*}\widetilde r_{n_*}=0$, 
 $n$ satisfies $n\geq n_*(x)$ 
 if $x$ is in the oscillatory region of $P_n(x)$. 
{}From \eqref{Rek}, for $n\geq n_*$, 
\begin{equation}
 \Repa W_r \left(x,\xi=\frac{n}{N}\right) = 
  \int_0^{\xi_*=\frac{n_*}{N}}\rmd\xi
  \left(\widehat k^{(0)}(\xi) + \widetilde k^{(0)}(\xi)
   -\frac{1}{2}\log\widehat r_n-\frac{1}{2}\log\widetilde r_n\right),
\end{equation}
hence $\Repa W_r(x,\xi)$ does not depend on $\xi$. 
Thus, we can set $W_r(x,\xi)=W_r(x,1)$ for $\xi\geq n_*/N$. 
As seen in \eqref{P_nin}, the frequency of the oscillation 
of $P_n(x)$ is of order $N$. Therefore, 
if we take the average of this oscillation for $P_n^2(x)$, 
it simply gives 1/2. 
We thus obtain 
\begin{align}
 \psi_n^2(x) &\simeq 2\frac{\left\lvert x\right\rvert\sqrt{\widehat r_0}}
  {\sqrt{4\widehat r_n \widetilde r_n
  -(x^2-\widehat r_n - \widetilde r_n)^2}}
  \exp\left[N\Repa K_r(x,1)
  \right] .
\end{align}
Here, the branch of the square root is chosen to be real and positive. 
This expression is convenient for calculating $D_N$ 
because of the independence of $\xi$ at order $N$. 

\section{Effective potential and resolvent}\label{app:VeffandR}

In this section, we clarify the relation 
between the effective potential and the resolvent. 
We show that the first derivative of the effective potential 
can be identified with the resolvent. 
The coefficient of the resolvent in the double scaling limit 
can be determined by this relation. 

The relation between the effective potential and resolvent 
can be derived as follows. 
Because $P_N(x)=\bracket{\det(x-\Phi)}$, 
we have 
\begin{equation}
 \bracket{\det(x-\Phi)} 
  = \left(\frac{x^2}
  {(x^2-\widehat r_n - \widetilde r_n)^2
  -4\widehat r_n \widetilde r_n}\right)^\frac14
  \exp\left[\frac{N}{2}W_0\left(x,\frac{n}{N}\right)
  +\frac{1}{2}k_n^{(0)}
  \right] ,
\end{equation}
for $x$ outside the cut.
Differentiating both sides with respect to $x$, 
we obtain 
\begin{equation}
 \bracket{\frac{1}{N}\tr\frac{1}{x-\Phi}\det(x-\Phi)} 
  = \bracket{\det(x-\Phi)}
  \left(\frac{1}{2}\partial_x K(x) + \mathcal O(\frac{1}{N})\right) .
\end{equation}
In the large-$N$ limit, the expectation value 
on the left-hand side factorizes. 
Hence, we have 
\begin{subequations}\label{ResolventandEffectivePotential}
 \begin{align}
 R(x) &= \bracket{\frac{1}{N}\tr\frac{1}{x-\Phi}} \\
  &= \frac{1}{2}\partial_x K(x)
  = \frac{1}{2}\partial_x
  \int_0^1\rmd\xi
 \left(\widehat k^{(0)}(\xi) + \widetilde k^{(0)}(\xi)\right). 
 \end{align}
\end{subequations}
The resolvent inside the cut can be obtained 
by analytic continuation, 
and it is consistent with the choice of the double sign 
in the definition of 
$\widehat k^{(0)}(\xi)$ and $\widetilde k^{(0)}(\xi)$ 
given in \eqref{khatdef} and \eqref{ktildedef}. 
We can evaluate the integration 
in \eqref{InsideCut} by using this relation. 
In the case of two cuts, 
we have 
\begin{equation}
 \partial_x
 \left(\widehat k^{(0)}(\xi) + \widetilde k^{(0)}(\xi)\right)
 = \frac{2x}{\sqrt{(x^2-\widehat r_n-\widetilde r_n)^2
 -4\widehat r_n \widetilde r_n}} ,\label{DiffofResolvent}
\end{equation}
which is again understood as an analytic function. 
Thus, the eigenvalue density can be obtained as 
\begin{subequations}
 \begin{align}
  \rho(x) &= 
  \frac{i}{2\pi}\bracket{\frac{1}{N}\tr\frac{1}{x+i\epsilon-\Phi}}
  -\frac{i}{2\pi}\bracket{\frac{1}{N}\tr\frac{1}{x-i\epsilon-\Phi}} \\
  &= \frac{1}{\pi}\Repa\int_{\xi_*}^1\rmd\xi
  \frac{x}
  {\sqrt{4\widehat r(\xi) \widetilde r(\xi)
  -\left(x^2-\widehat r(\xi) -\widetilde r(\xi)\right)^2}} .
 \end{align}
\end{subequations}
Here, the square root is again defined as an analytic function. 
The sign of the square root is positive for positive $x$. 
Note that $\rho(x)$ is positive even for negative $x$ 
because the square root also becomes negative there. 
Thus, the integration in \eqref{InsideCut} gives 
the eigenvalue density. 

The relation between the effective potential and 
the resolvent \eqref{ResolventandEffectivePotential} 
can be checked by a direct computation near the critical point, 
where it can be expressed as 
\begin{subequations}
 \begin{align}
  \partial_x
  &\left(\widehat k^{(0)}(\xi) + \widetilde k^{(0)}(\xi)\right)
  \simeq -\frac{x}{b(\xi)\sqrt{a^2(\xi)-x^2}} , \\
  & a(\xi)^2 \simeq \frac{\alpha^2}{r_c}(g_c-g\xi)
  \qquad \qquad b^2(\xi) \simeq 4r_c .
 \end{align}
\end{subequations}
Here, $a = a(1)$ and $b= b(1)$ are 
identified with the endpoints of the cut.
Integrating this equation with respect to $x$, 
the effective potential near the critical point 
can be obtained as 
\begin{subequations}
 \begin{align}
  \frac{1}{2}V_\eff^{\prime (0)}(x) 
  &= -\frac{1}{2}\partial_x\int_0^1\rmd\xi
  \left[\widehat k(\xi) + \widetilde k(\xi)\right]
  \simeq - C'x\sqrt{a^2-x^2} , \\
  & C' = \frac{\sqrt{r_c}}{\alpha^2 g} \qquad\qquad
  a^2 = \frac{\alpha^2}{r_c}(g_c-g) ,
 \end{align}
\end{subequations} 
and can be identified with the resolvent.

\section{Relation between $r$ in one-cut phase and in two-cut phase}
\label{app:relation}

There is a relation between the behavior of $r$ in the one-cut phase 
and in the two-cut phase. 
Here, we use the notations in \eqref{DefF}
\begin{align}
 g\xi = F(\widehat r,\widetilde r)
 &= g_c - \widehat A (r_c-\widehat r) 
 - \widetilde A (r_c-\widetilde r) \notag\\
 &\quad - \widehat B (r_c -\widehat r)^2 - \widetilde B (r_c - \widetilde r)^2
 - C (r_c - \widehat r)(r_c - \widetilde r), \tag{\ref{DefF}} 
\end{align}
and from \eqref{FandXi} we consider the case where 
$A = \widehat A = \widetilde A$. 
In the two-cut phase, $\widehat r$ and $\widetilde r$ 
satisfy the condition $(r_c-\widehat r) + (r_c-\widetilde r) = 0$. 
From these conditions, $r$ near the critical point 
in the two-cut phase is expressed as 
\begin{equation}
 r = r_c 
  \pm \left(\widehat B + \widetilde B - C\right)^{-\frac12}
  \sqrt{g_c-g\xi}.
\end{equation}
The double sign indicates $+$ for $\widetilde r$ 
and $-$ for $\widehat r$, because $\widehat r$ vanishes for $\xi=0$. 
In the one-cut phase, $r = \widehat r = \widetilde r$, 
and then $r$ near the critical point is 
\begin{equation}
 r = r_c - \frac{1}{2A}(g_c-g\xi).
\end{equation}

$A$, $\widehat B$, $\widetilde B$, and $C$ can be 
expressed in terms of $F(\widehat r,\widetilde r)$ as  
\begin{subequations}\label{ABBC}
 \begin{align}
  A &= \left.\deriv{F(\widehat r,\widetilde r)}{\widehat r}
  \right\rvert_{\widehat r = \widetilde r = r_c}
  = \left.\deriv{F(\widehat r,\widetilde r)}{\widetilde r}
  \right\rvert_{\widehat r = \widetilde r = r_c}, \\
  \widehat B &= -\frac{1}{2}\left.\deriv{^2 F(\widehat r,\widetilde r)}
  {\widehat r ^2}
  \right\rvert_{\widehat r = \widetilde r = r_c}, \\
  \widetilde B &= -\frac{1}{2}\left.\deriv{^2 F(\widehat r,\widetilde r)}
  {\widetilde r ^2}
  \right\rvert_{\widehat r = \widetilde r = r_c}, \\
  C &= -\left.\deriv{^2 F(\widehat r,\widetilde r)}
  {\widehat r \partial \widetilde r}
  \right\rvert_{\widehat r = \widetilde r = r_c} .
 \end{align}
\end{subequations}
Here, we use the fact that $F(\widehat r,\widetilde r)$ originates 
from $\left[gV'(X)\right]_{n,n-1}$, 
and that it can be expressed in terms of $V(x)$. 
Because the potential we consider is even, 
$V'(x)$ has only the odd power of $x$ 
and can be expressed as $gV'(x) = xU(x^2)$. 
Using this, $F(\widehat r,\widetilde r)$ is expressed as 
\begin{equation}
 F(\widehat r,\widetilde r) = \oint\rmd z
  \left(z+\frac{\widehat r}{z}\right)
  U(\left(z+\frac{\widehat r}{z}\right)
  \left(z+\frac{\widetilde r}{z}\right)), \label{FandU}
\end{equation}
for even $n$, 
and whereas odd $n$ we should use $F(\widetilde r,\widehat r)$. 
The condition $\widehat A = \widetilde A$ is not trivial, 
but comes from \eqref{FandXi}. 
Using \eqref{FandU}, \eqref{FandXi} is expressed as 
\begin{equation}
 0=\left(\widehat r - \widetilde r\right)\oint\frac{\rmd z}{z}
  U(\left(z+\frac{\widehat r}{z}\right)
  \left(z+\frac{\widetilde r}{z}\right)) . 
\end{equation}
To satisfy this condition with $\widehat r \neq \widetilde r$, 
we need 
\begin{equation}
 0=G(\widehat r, \widetilde r) 
  = \oint\frac{\rmd z}{z}
  U(\left(z+\frac{\widehat r}{z}\right)
  \left(z+\frac{\widetilde r}{z}\right)) .\label{GandU}
\end{equation}

From \eqref{ABBC}, \eqref{FandU}, and \eqref{GandU}, 
we can see 
\begin{equation}
 \widehat B + \widetilde B - C = \frac{A}{2r_c} .
\end{equation}
This is the relation between $r$ in the one-cut phase 
and in the two-cut phase.


\begin{thebibliography}{99}

\bibitem{Brezin:1990rb}
E.~Brezin and V.~A.~Kazakov,
``Exactly Solvable Field Theories Of Closed Strings,''
Phys.\ Lett.\ B {\bf 236}, 144 (1990). \\ 
M.~R.~Douglas and S.~H.~Shenker,
``Strings In Less Than One-Dimension,''
Nucl.\ Phys.\ B {\bf 335}, 635 (1990). \\ 
D.~J.~Gross and A.~A.~Migdal,
``Nonperturbative Two-Dimensional Quantum Gravity,''
Phys.\ Rev.\ Lett.\  {\bf 64}, 127 (1990), \\ 
``A Nonperturbative Treatment Of Two-Dimensional Quantum Gravity,''
Nucl.\ Phys.\ B {\bf 340}, 333 (1990). 
 
\bibitem{David:1990sk}
F.~David,
``Phases Of The Large N Matrix Model And Nonperturbative Effects In 2-D
Gravity,''
Nucl.\ Phys.\ B {\bf 348}, 507 (1991), \\  
``Nonperturbative effects in matrix models and vacua of two-dimensional
gravity,''
Phys.\ Lett.\ B {\bf 302}, 403 (1993)
arXiv:hep-th/9212106. \\
B.~Eynard and J.~Zinn-Justin,
``Large order behavior of 2-D gravity coupled to d $<$ 1 matter,''
Phys.\ Lett.\ B {\bf 302}, 396 (1993)
arXiv:hep-th/9301004.

\bibitem{Knizhnik:1988ak}
V.~G.~Knizhnik, A.~M.~Polyakov and A.~B.~Zamolodchikov,
``Fractal Structure Of 2d-Quantum Gravity,''
Mod.\ Phys.\ Lett.\ A {\bf 3}, 819 (1988). \\
F.~David,
``Conformal Field Theories Coupled To 2-D Gravity In The Conformal Gauge,''
Mod.\ Phys.\ Lett.\ A {\bf 3}, 1651 (1988). \\
J.~Distler and H.~Kawai,
``Conformal Field Theory And 2-D Quantum Gravity Or Who's Afraid Of Joseph Liouville?,''
Nucl.\ Phys.\ B {\bf 321}, 509 (1989).

\bibitem{Hanada:2004im}
M.~Hanada, M.~Hayakawa, N.~Ishibashi, H.~Kawai, T.~Kuroki, Y.~Matsuo and T.~Tada,
``Loops versus matrices: The nonperturbative aspects of noncritical string,''
Prog.\ Theor.\ Phys.\  {\bf 112}, 131 (2004)
arXiv:hep-th/0405076.

\bibitem{Douglas:2003up}
M.~R.~Douglas, I.~R.~Klebanov, D.~Kutasov, J.~Maldacena, E.~Martinec and N.~Seiberg,
``A new hat for the c = 1 matrix model,''
arXiv:hep-th/0307195.

\bibitem{Klebanov:2003wg}
I.~R.~Klebanov, J.~Maldacena and N.~Seiberg,
``Unitary and complex matrix models as 1-d type 0 strings,''
arXiv:hep-th/0309168.

\bibitem{Distler:1989nt}
J.~Distler, Z.~Hlousek and H.~Kawai,
``Superliouville Theory As A Two-Dimensional, Superconformal Supergravity Theory,''
Int.\ J.\ Mod.\ Phys.\ A {\bf 5}, 391 (1990).

\bibitem{Gross:1980he}
D.~J.~Gross and E.~Witten,
``Possible Third Order Phase Transition In The Large N Lattice Gauge Theory,''
Phys.\ Rev.\ D {\bf 21}, 446 (1980).

\bibitem{Zamolodchikov:2001ah}
A.~B.~Zamolodchikov and A.~B.~Zamolodchikov,
``Liouville field theory on a pseudosphere,''
arXiv:hep-th/0101152.

\bibitem{Fateev:2000ik}
V.~Fateev, A.~B.~Zamolodchikov and A.~B.~Zamolodchikov,
``Boundary Liouville field theory. I: Boundary state and boundary  two-point function,''
arXiv:hep-th/0001012. \\
J.~Teschner,
``Remarks on Liouville theory with boundary,''
arXiv:hep-th/0009138.\\
B.~Ponsot and J.~Teschner,
``Boundary Liouville field theory: Boundary three point function,''
Nucl.\ Phys.\ B {\bf 622}, 309 (2002)
arXiv:hep-th/0110244.

\bibitem{Demeterfi:1990gb}
K.~Demeterfi, N.~Deo, S.~Jain and C.~I.~Tan,
``Multiband Structure And Critical Behavior Of Matrix Models,''
Phys.\ Rev.\ D {\bf 42}, 4105 (1990).

\bibitem{DiFrancesco:1993nw}
P.~Di Francesco, P.~H.~Ginsparg and J.~Zinn-Justin,
``2-D Gravity and random matrices,''
arXiv:hep-th/9306153.



\end{thebibliography}
\end{document}